\documentclass[10pt,twocolumn,preprintnumbers,superscriptaddress,nofootinbib,aps,prd]{revtex4-2}

\usepackage{graphicx}
\usepackage{amsmath}
\usepackage{srcltx}
\usepackage[utf8]{inputenc}
\usepackage[colorlinks]{hyperref}
\usepackage{url}
\usepackage{times}
\usepackage{amssymb}

\newcommand{\ord}[1]{\mathcal{O}{(#1)}}
\newcommand{\beq}{\begin{equation}}
	\newcommand{\eeq}{\end{equation}}
\newcommand{\bea}{\begin{eqnarray}}
	\newcommand{\eea}{\end{eqnarray}}

\newcommand{\appropto}{\mathrel{\vcenter{
			\offinterlineskip\halign{\hfil$##$\cr
				\propto\cr\noalign{\kern2pt}\sim\cr\noalign{\kern-2pt}}}}}

\begin{document}
	
	\pagestyle{plain}
	
	\title{Electron-Ion Collider as a Discovery Tool for Invisible Dark Bosons}

\author{Hooman Davoudiasl}
\email{hooman@bnl.gov}

\author{Hongkai Liu} 
\email{hliu6@bnl.gov}

\affiliation{High Energy Theory Group, Physics Department\\ Brookhaven National Laboratory,
	Upton, NY 11973, USA}
	
	
	\begin{abstract}

We illustrate how the future Electron-Ion Collider (EIC) can be used to discover dark bosons with masses in the $\sim$ (10~MeV -- 10~GeV) regime, having a wide range of properties.  We only require that the dark bosons have a non-negligible weak coupling to electrons and decay with $\ord{1}$ branching fraction into invisible final states.  Our signal selection takes advantage of the excellent electron beam kinematic measurements and the capability to tag incoherent scattering, as envisioned at the EIC.  This makes the EIC a powerful tool for uncovering potential dark sector forces, for a variety of possibilities.

	\end{abstract}
	\maketitle

\section{Introduction}

While important fundamental questions remain open in particle physics and cosmology, there is currently no obvious hint where the answers may come from.  Of these questions, the origin of neutrino masses and the nature of dark matter (DM) are two of the most pressing that we face today.  Even though the evidence for these puzzles is not in doubt, the problem with either one -- and some other fundamental mysteries -- is that they can be resolved in a multitude of ways and over very wide ranges of parameters.  This circumstance compels us to use every tool and all available methods to eliminate the allowed parameter space and, with a little help from luck, perhaps find a glimpse of the answers. 

The Electron-Ion Collider (EIC), to be built over the coming years at the Brookhaven National Laboratory, aims to investigate the detailed structure of hadronic matter~\cite{AbdulKhalek:2021gbh}.  However, the capabilities of this new facility and its experiments can also lend themselves to searching for a variety of new phenomena that lie outside the Standard Model (SM)~\cite{Gonderinger:2010yn,Boughezal:2020uwq,Liu:2021lan,Cirigliano:2021img,Davoudiasl:2021mjy,Yan:2021htf,Li:2021uww,Batell:2022ogj,Zhang:2022zuz,Yan:2022npz,Boughezal:2022pmb,Davoudiasl:2023pkq,Balkin:2023gya,Davoudiasl:2024vje,Wang:2024zns,Wen:2024cfu,Gao:2024rgl,Du:2024sjt,Deng:2025hio}.  This possibility has gained significant attention in recent years and is worth further investigation.  Given the parameters of the EIC, new physics at low mass scales with weak coupling to the SM would be a natural target for such experimental searches.  These types of models have also been a focus of much attention over the last decade or so, and have been invoked in addressing various open questions~\cite{Alexander:2016aln}.  In particular, such low scale physics may have its own ``dark" or ``hidden" sector that includes DM, as well as a number of other particles and forces that only indirectly interact with the SM, {\it i.e.} the  ``visible" world.

Motivated by the above considerations, in this work we will consider the detection of dark bosons, representing broad classes of models, at the EIC.  We will only assume a very minimal set of requirements for the dark boson properties, mainly that: {\it (i)} it have a non-negligible weak coupling to electrons, and {\it (ii)} its decay have $\ord{1}$ branching fraction into invisible states.  The first requirement stems from our focus on a production mechanism where the boson is emitted from the electron beam, in low momentum transfer coherent electromagnetic scattering from large atomic number $Z$ ions,  like gold.  This process would have a $Z^2$ enhanced rate and can be quite important for producing GeV scale bosons emitted from the electron beam.  The second requirement originates from our signal selection criteria that depend on the kinematics of beam electron upon dark boson emission.  By focusing on invisible final states, we could avoid large SM backgrounds and have a clean signal.  We note that the above conditions could be realized in a variety of models, including gauged $B-L$ (with $B$ and $L$ baryon and lepton numbers, respectively) \cite{Mohapatra:1974hk,Mohapatra:1974gc,Senjanovic:1975rk}, gauged $L_e-L_i$ models with $i=\mu,\tau$ \cite{He:1990pn}, dark $Z$ models characterized by a dark vector mixing with the SM $Z$ boson \cite{Davoudiasl:2012ag}; in these models neutrinos contribute significant branching fractions to the decay of the new vector boson.  In general, one could also consider models where dark sector states, or neutrinos, constitute the main decay branching fraction of a dark boson, which could be a scalar or a vector.     
\begin{figure}[h!]
	\centering
	{\includegraphics[width=0.45\linewidth]{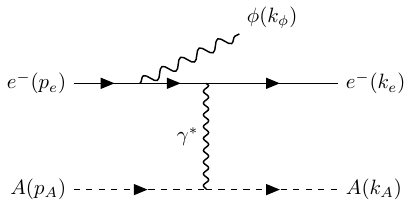}}
        {\includegraphics[width=0.45\linewidth]{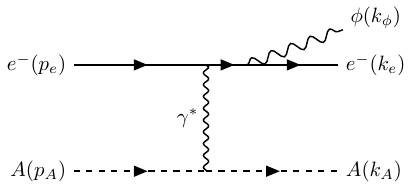}}
	\caption{Feynman diagrams of the $\phi$ production via  bremsstrahlung from the electron. We include both ISR (left panel) and FSR (right panel).
	}	
	\label{fig:brem}
\end{figure}

\section{Basic Models}

We will consider two cases: a scalar and a vector boson, both denoted by $\phi$.  These will suffice to outline the broad classes of models that could be investigated at the EIC,  using our proposed methodology.
We consider the following generic interactions:
\bea
\mathcal{L}_S&=&g_S^e\, \phi \bar e e + g_S^\chi\, \phi \bar \chi \chi\,,\nonumber\\
\mathcal{L}_V&=&g_V^e\, \phi_\mu \bar e \gamma^\mu e + g_V^\chi\, \phi_\mu \bar \chi \gamma^\mu \chi\,,
\label{phi}
\eea
where $g_{S(V)}^e$ and $g_{S(V)}^\chi$ are scalar (vector) couplings to electrons and $\chi$, respectively.
The new boson $\phi$ can be produced via the electron bremsstrahlung as shown in Fig.~\ref{fig:brem}.
We assume $\chi$ is an invisible particle (SM neutrino or dark-sector particle) at the detector length scale 
 and its mass $m_\chi < m_\phi/2$, so that $\phi \to \bar \chi \chi$ is allowed on-shell. In principle, $\chi$ could be an SM particle or decay back to the SM particles, leading to a displaced vertex in the detector~\cite{Davoudiasl:2023pkq}. The combination of a displaced vertex and a depleted electron energy signal could potentially enhance the sensitivity. However, these scenarios require a more detailed analysis, which we leave for future work.

\section{Analysis}  

\begin{figure}
	\centering
	\includegraphics[width=0.9\columnwidth]{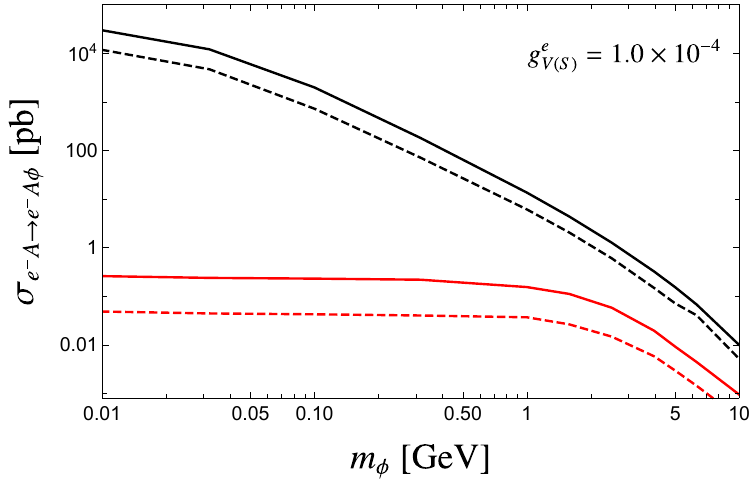}
	\caption{The production cross sections of $\phi$ via electron bremsstrahlung. The solid (dashed) black line shows the production cross section of the new vector (scalar) boson. The red lines show the corresponding production cross sections after applying the cuts in Table~\ref{tab:cuts}.}
	\label{fig:xs}
\end{figure}

The EIC can precisely measure the recoil electron and is capable of distinguishing between coherent scattering, where the nucleus remains intact, and incoherent scattering, where the nucleus breaks up, including through hard scattering and jet emission.
In this work, we consider the coherent production of $\phi$, which benefits from a large $Z^2$ enhancement for heavy nuclei such as gold. 
In the limit $m_e \ll m_\phi\ll m_A \ll \sqrt{s}$, the transferred momentum to the
nucleus $Q_A^2 \sim m_\phi^4 m_A^2/s^2$, where $m_A$ denotes the ion mass and $\sqrt{s}$ is the center of mass energy. The required $Q_A^2$ increases rapidly with $m_\phi$. The ion-photon interaction vertex in the coherent regime is $i V^\mu(Q_A^2, p_A, k_A) = i e Z F(Q_A) (p_A+k_A)^\mu $.
The nuclear form factor $F(Q_A)$ strongly suppresses contributions from $\sqrt{Q_A^2}$  much larger than the inverse of the nucleus size, $r_A^{-1} \sim (A^{1/3}\,\text{fm})^{-1}$, where $A$ is the mass number of the nucleus. By assuming an electron beam energy 18 GeV and a 100 GeV per nucleon ion beam \cite{AbdulKhalek:2021gbh}, we can estimate the maximum boson mass for coherent production $(m_\phi)_{\rm max} \sim 20~{\rm GeV} (197/A)^{1/6}$. 
The production cross sections drop sharply with increasing $m_\phi$ as shown by the solid curves in Fig.~\ref{fig:xs}. The amplitude squared and phase space integration are discussed in detail in~\cite{Balkin:2023gya,Davoudiasl:2023pkq}. To account for the nuclear charge distribution at low momentum transfer, we use the Helm form factor~\cite{Helm:1956zz} 
\begin{align}
    F(Q_A) 
=   \frac{3j_1(Q_A R_1)}{Q_A R_1}
    \text{exp}\left[-\frac{(Q_A \rho)^2}{2}\right] \, ,
\end{align}
where $j_1$ is the first spherical Bessel function of the first kind, $\rho=0.9$~fm and $R_1\simeq 1.1$ fm $A^{1/3}$.
We find similar kinematics for scalar and vector dark boson $\phi$. Therefore, in the following, we present only the results for the vector boson. Our final projections for the coupling $g^e_\phi$ can be obtained for the scalar case by simple scaling of the cross section $\sigma_\phi$, via the relation $\sqrt{\sigma_\phi} \propto g^e_\phi$.

\begin{figure}
	\centering
	\includegraphics[width=0.9\columnwidth]{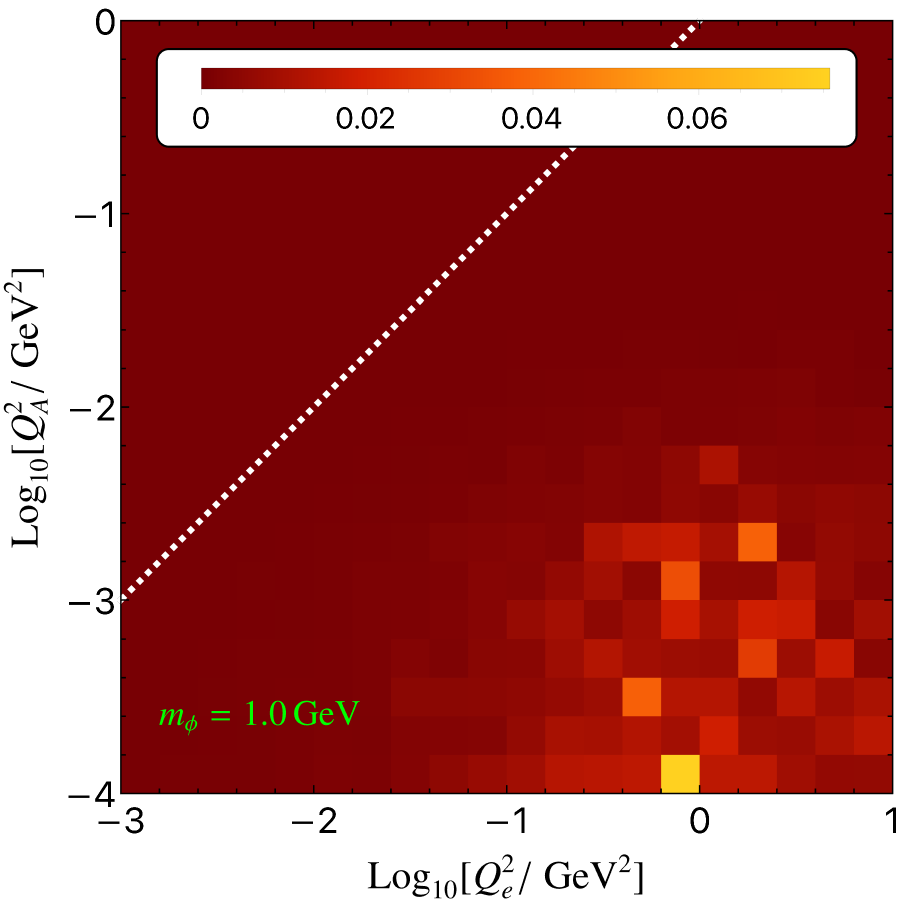}
	\caption{The correlation between $Q_e^2$ and $Q_A^2$ in the presence of $\phi$ with a mass of 1.0 GeV.  The dashed white line represents the SM background expectation.}
	\label{fig:Q2}
\end{figure}

\begin{figure}
	\centering
	\includegraphics[width=0.9\columnwidth]{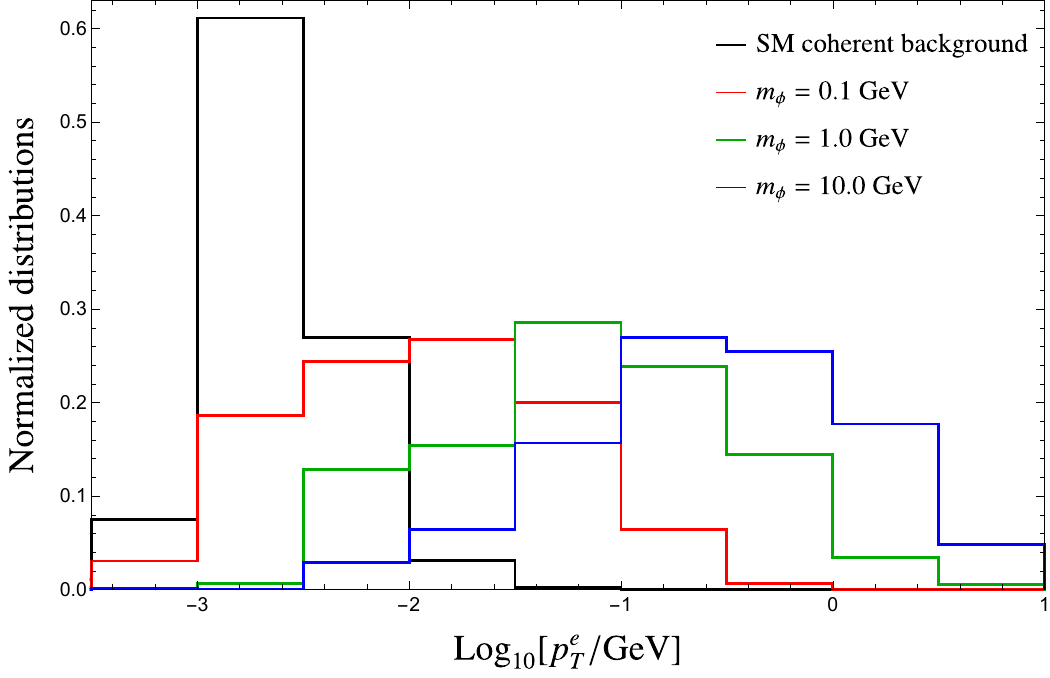}
    \includegraphics[width=0.9\columnwidth]{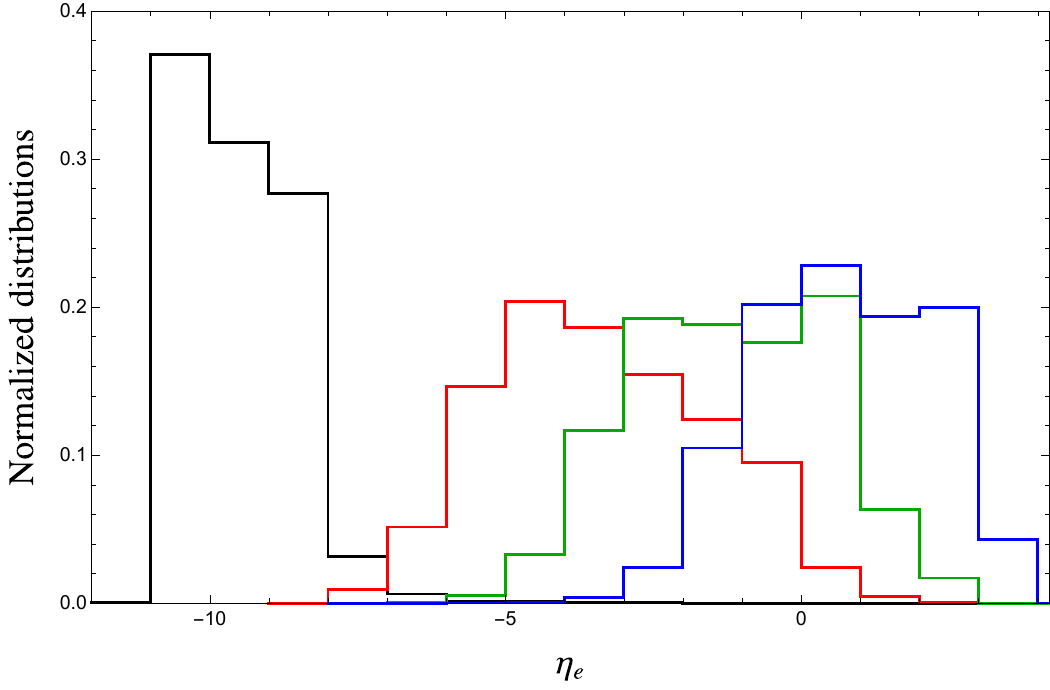}
	\caption{The recoil electron distributions of transverse momentum (upper panel) and pseudorapidity (lower panel).}
	\label{fig:dis}
\end{figure}

The new boson $\phi$ with a mass of $\mathcal{O}(\rm GeV)$ carries away most of the electron beam energy, leaving a soft electron in the final state. Therefore, the transferred momentum on the electron side~($Q_e^2$) is significantly larger than that on the ion side~($Q_A^2$), as illustrated in Fig.~\ref{fig:Q2}.
This behavior differs from the SM coherent scattering, where both the electron and ion experience low and comparable momentum transfer.   
The distributions of transverse momentum $p_T^e$ and pseudorapidity $\eta_e$ of the recoil electrons are shown by the colored lines in Fig.~\ref{fig:dis}, for three representative $\phi$ masses. 
As the virtual photon exchanged between the ion and electron becomes harder for heavier $\phi$, the recoil electron tends to shift towards the central region and can even move in the forward direction in the production of $\mathcal{O}$(GeV) scale $\phi$. 

One of the main backgrounds to the above $\phi$ production process is the SM bremsstrahlung $e^- A \to e^- A \,\gamma$, where the photon is missed. In contrast to the massive vector dark boson production in the signal, the SM bremsstrahlung background typically features soft and collinear photons. The electron distributions are depicted by the black lines in Fig.~\ref{fig:dis}.
Therefore, as summarized in Table~\ref{tab:cuts}, we impose mass-dependent selection cuts requiring central electrons with large transverse momenta to suppress the SM bremsstrahlung background.
\begin{table*}[t]
\centering
\small
\begin{tabular}{|c|c|c|c|c|c|c|c|}
\toprule
$m_\phi$ [GeV] & $Q^2_{e,\min}$ [GeV$^2$] & $p_{T,e}^{\min}$ [GeV] &
$\eta_e^{\min}$ & $\eta_e^{\max}$ & $E_e^{\max}$ [GeV] &
$\sigma_{\gamma}^{\rm eff}$ [pb] & $(1-\epsilon_{\rm veto})~A~\sigma_{\rm DIS}^{\rm eff}$ [pb] \\
0.010 & $ 10^{-0.7}$ & 1.1 & -3.5 & -1.5 & 10.0 & 2.06 & 0.0310 \\
0.032 & $ 10^{-0.7}$ & 1.1 & -3.5 & -1.5 & 10.0 & 2.06 & 0.0310 \\
0.100 & $ 10^{-0.7}$ & 1.1 & -3.5 & -1.5 & 10.0 & 2.06 & 0.0310 \\
0.316 & $ 10^{-0.7}$ & 1.1 & -3.5 & -1.5 & 10.0 & 2.06 & 0.0310 \\
1.000 & $ 10^{0.2}$ & 1.3 & -3.5 & 2.0 & 10.0 & 1.93 & 0.0520 \\
1.585 & $ 10^{0.5}$ & 1.3 & -3.0 & 2.0 & 10.0 & 1.93 & 0.0520 \\
2.512 & $ 10^{0.8}$ & 1.3 & -2.5 & 2.5 & 10.0 & 1.54 & 0.0520 \\
3.981 & $ 10^{1.2}$ & 1.3 & -2.5 & 1.0 & 10.0 & 0.70 & 0.0330 \\
5.000 & $ 10^{1.4}$ & 1.3 & -3.0 & 1.0 & 10.0 & 0.43 & 0.0204 \\
6.310 & $ 10^{1.5}$ & 1.3 & -2.0 & 3.5 & 10.0 & 0.33 & 0.0155 \\
10.000 & $ 10^{1.5}$ & 1.3 & -1.5 & 3.0 & 10.0 & 0.33 & 0.0155 \\
\end{tabular}
\caption{Cuts used for the search of $\phi$ boson at different masses, and the corresponding effective background cross sections after cuts;  $\epsilon=10^{-6}$ and $\epsilon_{\rm veto}=95\%$ are assumed.}
\label{tab:cuts}
\end{table*}
The signal cross sections after applying the cuts are given by the red lines in Fig.~\ref{fig:xs}, which as expected from the preceding discussion are significantly more suppressed in the low $m_\phi$ regime.   
Since we select recoil electrons with energy less than 10 GeV, the photon energy in the SM bremsstrahlung is mostly larger than 5~GeV. We assume that the inefficiency for 5 GeV central photons is $10^{-8}\leq \epsilon \leq 10^{-5}$~\cite{Maeda:2014pga,KOTO:2025gvq}. In our analysis, the chance of missing such energetic photons is conservatively taken to be 100\%, when $|\eta_\gamma| > 3.5$. The effective SM bremsstrahlung cross sections after applying the cuts are given in Table~\ref{tab:cuts} assuming inefficiency $\epsilon = 10^{-6}$.

Another main background is the SM deep inelastic scattering (DIS) $e^- A \to e^- X j$, where $X$ denotes nuclear hadronic debris, and the jet $j$ is missed. Since the EIC detector operational capabilities are not established and an envisioned second detector has not been defined yet, we only focus on the gross features of the signal, such as the expected total hadronic energy in the central detector. For this purpose, we show the results at the parton level and use \texttt{MadGraph5\_aMC@NLO}~\cite{Alwall:2011uj} to simulate the DIS process $e p \to e X j$.

Our simulations show that, after applying the selection cuts, the jet produced in the DIS process remains central($|\eta_j|<3.5$ ), with a minimum energy of 8 GeV. Therefore, similar to the SM bremsstrahlung, we expect significant activity in the central detector due to the jet emission.  Additionally, the DIS background can be reduced significantly by the veto of incoherent scattering at the Zero Degree Calorimeter \cite{AbdulKhalek:2021gbh} in the far-forward direction with an efficiency $\epsilon_{\rm veto}\gtrsim 95\%$~\cite{Chang:2021jnu,Aschenauer:2025mku}.  Hence, we assume that the DIS background is suppressed by an inefficiency $10^{-8}\leq \epsilon \leq 10^{-5}$ for missing such an energetic and central jet, similar to the coherent background discussed earlier.  
The effective DIS background cross sections are given in Table~\ref{tab:cuts}, as well. We note that the DIS background can be enhanced by a factor of $A$ to account for the corresponding nucleon luminosity. The DIS background is quite a bit smaller than that of the bremsstrahlung process and is hence ignored. The cross section of the SM irreducible background $e^- A \to e^- A \nu \bar\nu$ mediated by the massive weak gauge bosons is found to be negligible compared to the previous two background processes.

\begin{figure}
	\centering
	\includegraphics[width=0.9\columnwidth]{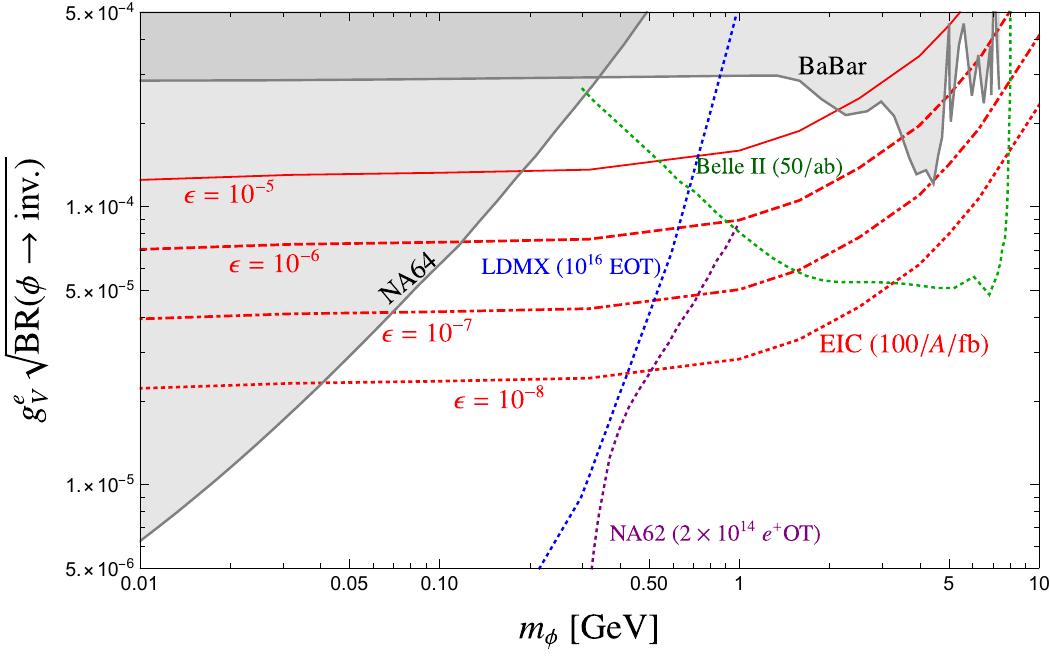}
	\caption{The EIC projections for the electron coupling $g_V^e$ of an invisibly decaying dark vector $\phi$,  with invisible branching fraction $\text{BR} (\phi\to \text{inv.})$. The $2~\sigma$ bounds corresponding to $10^{-8}\leq \epsilon \leq 10^{-5}$, are given by the red curves.  An integrated luminosity of $100/A$ fb$^{-1}$ has been assumed for $Z=79$, corresponding to gold ions.  The current constraints are shaded in gray~\cite{BaBar:2017tiz,NA64:2023wbi}. The Belle-II~\cite{Essig:2013vha,Belle-II:2018jsg}, LDMX~\cite{Izaguirre:2014bca,Izaguirre:2015yja}, and NA62~\cite{Arias-Aragon:2025qod} projections are overlaid for comparison.}
	\label{fig:gVe-limit}
\end{figure}

The signal cross sections after the cuts are shown by the red curves in Fig.~\ref{fig:xs}.
To estimate the reach of the EIC, we require ${\cal S}/\sqrt{\cal B} = 2$ with $100/A$ fb$^{-1}$ integrated $eA$ luminosity (assuming ultimate design luminosity). 
 Here, ${\cal S}$ and ${\cal B}$ are the numbers of signal and background events, respectively. The 2 $\sigma$ projections for the reach of the EIC are shown by the four red lines in Fig.~\ref{fig:gVe-limit}, corresponding to four different choices of the detector inefficiency $\epsilon$, ranging from $10^{-5}$ to $10^{-8}$.
The most stringent bounds on the invisibly decaying dark vector boson are from the mono-photon search at BaBar~\cite{BaBar:2017tiz} and the missing momentum search at NA64~\cite{NA64:2023wbi}. 

As can be seen from Fig.\ref{fig:gVe-limit}, the EIC has the potential to probe a wide range of open parameter space for dark bosons, as long as they have a non-negligible branching fraction into modes that are invisible on detector transit time scales.  In particular, for such bosons over much of the mass range $\sim$ (0.3 -- 10) GeV, the projected EIC constraints on the associated models can be better than those of current or envisioned experiments in the future.     

\section{Summary} 
In this work, we have considered the potential for the EIC to uncover the signals of dark bosons that couple weakly to electrons and decay with significant invisible branching fraction.  These bosons, vectors or scalars, can originate from a broad class of models.  The invisible final states can be neutrinos, or they may be light states from a dark sector that escape the detector due to their sufficiently long lifetimes. 

We examine a search strategy that leverages the excellent electron detection capabilities of the future EIC experiments, where emission of the invisible boson leads to considerable depletion of the electron beam energy and imparts a significant transverse momentum to it.  The production process we consider is based on coherent scattering of the electron from a high $Z$ ion, like gold, and leads to a large cross section.  The intact ion in this process also provides another handle on signal identification.  The main background, from the analogue process with the emission of a hard GeV scale photon (coherent) or jet (DIS), can be suppressed assuming realistic detection efficiencies.  The projected parameter space that can be probed for these models is broad and can extend beyond those accessed in the future, by current or proposed experiments.  We conclude that the EIC can play a significant role in the experimental investigations of dark sector forces in a wide variety of interesting models.   

The digital data associated with this work can be found as ancillary files with the arXiv submission~\cite{datanote}.

	~
	\begin{acknowledgments}
		We thank A. Jentsch for helpful discussions regarding experimental issues.  We also thank Rojae Mighty for pointing out numerical errors in the previous version of our results, which we have revised in the current manuscript.  This work is supported by the US Department of Energy under Grant Contract DE-SC0012704.
	\end{acknowledgments}

        \bibliography{LLP.bib}

\end{document}